\documentclass[fleqn,10pt]{wlscirep}
\usepackage[utf8]{inputenc}
\usepackage[T1]{fontenc}
\title{Revised Magnetic Structure and Tricritical Behavior of the CMR Compound NaCr$_2$O$_4$ Investigated with High Resolution Neutron Diffraction and $\mu^+$SR}

\author[1,*]{E.~Nocerino}
\author[2]{O.~K.~Forslund}
\author[3]{H.~Sakurai}
\author[4]{A.~Hoshikawa}
\author[1]{N. Matsubara}
\author[5]{D.~Andreica}
\author[6]{A.~Zubayer}
\author[7]{F.~Mazza}
\author[8]{J.-C.~Orain}
\author[9]{T.~Saito}
\author[10,11]{J.~Sugiyama}
\author[12]{I. Umegaki}
\author[2]{Y.~Sassa}
\author[1,$\dagger$]{M.~M{\aa}nsson}

\affil[1]{KTH Royal Institute of Technology, Department of Applied Physics, Alba Nova University Center, Stockholm, SE-114 21, Sweden}
\affil[2]{Chalmers University of Technology, Department of Physics, G$\ddot{o}$teborg, SE-412 96, Sweden}
\affil[3]{National Institute for Materials Science, Namiki, Tsukuba, Ibaraki 305-0044, Japan}
\affil[4]{Frontier Research Center for Applied Atomic Sciences, Ibaraki University, 162-1 Shirakata, Tokai, Ibaraki 319-1106, Japan}
\affil[5]{Faculty of Physics, Babes-Bolyai University, 3400 Cluj-Napoca, Romania}
\affil[6]{Department of Physics, Chemistry and Biology (IFM), Linköping University, SE-581 83 Linköping, Sweden}
\affil[7]{Insitute of Solid State Physics, TU Wien, Wiedner Haupstraße 8-10, AT-1040 Wien (Austria)}
\affil[8]{Laboratory for Muon Spin Spectroscopy, Paul Scherrer Institute, CH-5232, Villigen, PSI, Switzerland}
\affil[9]{Institute of Materials Structure Science, High Energy Accelerator Research Organization, 203-1 Shirakata, Tokai, Ibaraki 319-1107, Japan}
\affil[10]{Neutron Science and Technology Center, Comprehensive Research Organization for Science and Society (CROSS), Tokai, Ibaraki 319-1106, Japan}
\affil[11]{Advanced Science Research Center, Japan Atomic Energy Agency, Tokai, Ibaraki 319-1195, Japan}
\affil[12]{Muon Science Laboratory, Institute of Materials Structure Science, KEK, Tokai, Ibaraki 319-1106, Japan}

\begin{abstract}
The mixed valence Cr compound NaCr$_2$O$_4$, synthesized using a high-pressure technique, offers a unique playground for investigating unconventional physical properties in condensed matter. In the present study, muon spin rotation/relaxation ($\mu^+$SR) and high-resolution neutron powder diffraction (NPD) measurements were carried out to clarify the true magnetic ground state of this interesting compound. Our detailed study brings new insight, allowing us to confirm the existence of a commensurate antiferromagnetic order (C-AFM) and to extract its ordered Cr moment $\mu^{\rm C}_{\rm Cr}=(4.30\pm0.01)\mu_B$. Such a value of the ordered moment is in fact compatible with the existence of high-spin Cr sites. Further, the value of the canting angle of the Cr spin axial vector is refined as $\theta_{\rm c}=(8.8\pm0.5)^{\circ}$. Employing high-quality samples in combination with time-of-flight NPD, a novel magnetic supercell was also revealed. Such supercell display an incommensurate (IC)-AFM propagation vector (0~0~${\textstyle \frac{1}{2}-}\delta$), having an ordered moment $\mu^{\rm IC}_{\rm Cr}=(2.20\pm0.03)\mu_B$. It is suggested that the C-AFM and IC-AFM modulations are due to itinerant and localized contributions to the magnetic moment, respectively. Finally, the direct measurement of the magnetic order parameter provided a value of the critical exponent $\beta = 0.245 \approx \frac{1}{4}$, suggesting a non conventional critical behavior for the magnetic phase transition in NaCr$_2$O$_4$.
\end{abstract}
\begin{document}

\flushbottom
\maketitle
%
%
\thispagestyle{empty}


\section*{Introduction}

\begin{figure}[ht]
  \begin{center}
    \includegraphics[scale=0.7]{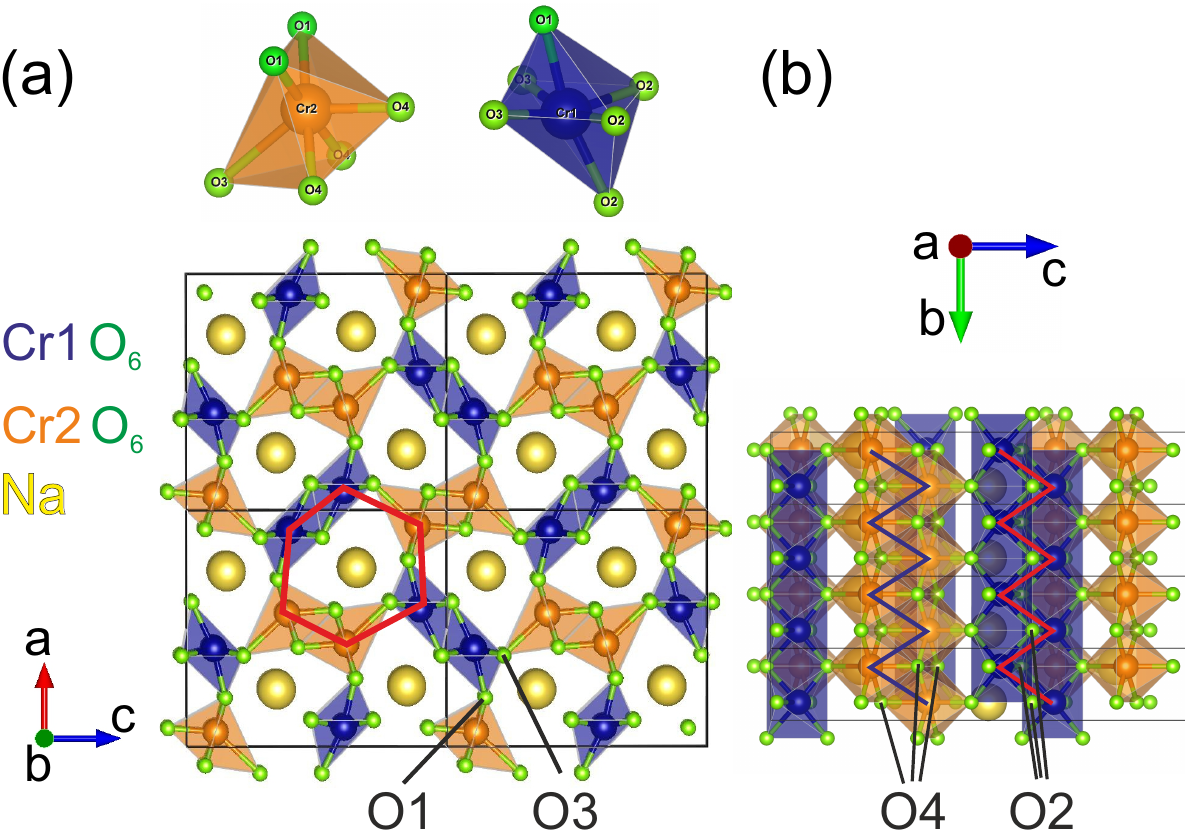}
  \end{center}
  \caption{Crystal structure of NaCr$_2$O$_4$. (a) $ac-$plane orientation of the unit cell; the hexagonal channel that hosts the Na$^+$ ions is highlighted in red. A detailed picture of the distorted octahedra for both Cr1 and Cr2 sites are also shown. The crystallographic sites for the O atoms are explicitly displayed to evidence the uneven distribution of the Cr-O distances. (b) $bc-$plane orientation of the unit cell; the zig-zag chains along the $b-$axis are highlighted by the solid lines.}
  \label{struct}
\end{figure}

Mixed valence transition metal oxides constitute a very attractive scientific playground as they offer a vast range of diverse physical properties, such as unconventional superconductivity, metal-insulator transitions, colossal magneto-resistance and exotic magnetic ground states \cite{adler1968mechanisms,toriyama2011peierls,NocerinoLTO,bednorz1986possible,urushibara1995insulator,matsubara2020neutron, nozaki2010incommensurate}.
Among them, mixed valence state Cr oxides are rather rare because of the intrinsic stability of Cr$^{3+}$ ions in octahedral coordination with O atoms. Here the half-filled t$_2g$ states of the $3d^3$ Cr$^{3+}$ orbitals tend to not coexist with Cr$^{4+}$ ions, which would instead form a less-than-half filled $3d^2$ state.
Due to the plethora of its unusual physical properties, a lively interest surrounds the compound NaCr$_2$O$_4$. Here, Cr has formal valence Cr$^{3.5+}$, given by the mix in equal parts of Cr$^{3+}$ and Cr$^{4+}$ valence states \cite{taguchi2017unusual}, which can only be achieved thanks to the high pressure synthesis technique employed to stabilize this material \cite{sakurai2012unconventional}. The high pressure indeed forces a complex crystal-field symmetry and orbital ordering for the Cr ions in NaCr$_2$O$_4$ \cite{toriyama2014electronic}, which crystallizes in a calcium ferrite-type structure with space group \textit{Pnma}. here, the Cr atoms occupy the two distinct crystallographic positions Cr1 and Cr2, octahedrally coordinated by O atoms (Fig.~\ref{struct}). 
The CrO$_6$ octahedra associated to the two sites are tilted with respect to the each other and distorted, with $\widehat{O-Cr-O} \neq 90^{\circ}$ angles and Cr-O distances that are equal to each other on one side of the octahedron and different of the other side [Fig.~\ref{struct}(a)], leading to four crystallographically different oxygen sites. The edge sharing connections between the equal Cr-O sides of the octahedra, form zig-zag Cr chains along the $b-$axis, via the Cr1-O2-Cr1 and the Cr2-O4-Cr2 bonds. The corner sharing connections between the non-equal Cr-O sides, instead, form rutile slabs in the $ac-$plane via the Cr1-O1-Cr2 (along $a$) and the Cr1-O3-Cr2 (along $c-$) bonds [Fig.~\ref{struct}(a,b)].

The difference between the two Cr sites lays in their orbital structure, since different distortions in the CrO$_6$ octahedra lift the degeneracy on the Cr $3d$ orbitals in different ways. In particular, following the labelling of the current work, for Cr1 the $d_{xy}$ orbital is almost fully occupied with a localized nature, while the $d_{yz\pm xz}$ have an itinerant nature. Vice versa, for Cr2 the $d_{xy}$ and $d_{yz+xz}$ are almost fully occupied while the $d_{yz-xz}$ has an itinerant nature \cite{toriyama2014electronic}. Such a combination of orbital filling leads to a dual itinerant-localized character for the Cr electrons. Such CrO$_6$ network, designs honeycomb-like distorted hexagonal one-dimensional channels (along $b-$axis), in which the Na$^+$ ions are located [Fig.~\ref{struct}(a)].

NaCr$_2$O$_4$ exhibits an antiferromagnetic (AFM) transition $T_{\rm N}=125$~K and, below this temperature, it manifests colossal magneto-resistance (CMR) \cite{sakurai} as well. This is already surprising in itself since CMR is usually a physical property related to manganites \cite{tokura2000colossal} but, in addition, the CMR of NaCr$_2$O$_4$ is also anomalous. In conventional CMR systems, the presence of an applied magnetic field causes a dramatic reduction of the electrical resistivity in a limited temperature interval around a ferromagnetic transition. This effect causes hysteresis in the magnetic and transport measurements, and the system is still metallic even in absence of magnetic field. Under zero magnetic field, NaCr$_2$O$_4$ behaves as an insulator with its electrical resistivity diverging while cooling. In the presence of an applied magnetic field, the resistivity is dramatically reduced (i.e. the CMR effect), but this effect is not limited to the region of temperatures around the AFM transition. On the contrary, it is more and more enhanced as the temperature is lowered, without any thermal or field hysteresis.
The mechanism underlying conventional CMR in manganites is explained as a magnetic field-induced suppression of the thermal fluctuation of the parallel moments, causing the CMR to disappear at temperatures below the FM transition \cite{furukawa1995temperature}. Since the CMR in NaCr$_2$O$_4$ shows a behavior that is the exact opposite of the conventional case, the underlying mechanism must be entirely new.

Beyond the anomalous CMR, NaCr$_2$O$_4$ was also found to exhibit an unusual coexistence of positive and negative charge transfer states (via XAS measurements \cite{taguchi2017unusual}). Here, the charge frustration induced by the highly degenerate charge ordering configurations of the Cr$^{4+}$ $3d^2$ electronic state, is removed by establishing a $3d^3$ \underline{L} state (as suggested by $^{53}$Cr-NMR measurements \cite{takeda2013magnetic}), in which Cr ions share a hole through the ligand oxygen. Calculations of the electronic structure showed that such holes are unevenly distributed and are found preferentially at the corner sharing oxygen sites O1 and O3 \cite{toriyama2014electronic} (fig. \ref{struct}(a)).

Concerning the magnetic properties of NaCr$_2$O$_4$, magnetic susceptibility measurements displayed a large positive value for the Curie-Weiss temperature, indicative of dominant ferromagnetic (FM) interactions. However, the AFM character of the susceptibility (showing the typical AFM cusp), suggested non negligible AFM interactions \cite{sakurai2012unconventional}. Remarkably, a dramatic increase in the magnetization with the application of a relatively small external field of 3.5 T was also observed in bulk magnetic measurements \cite{sakurai2012unconventional}. This behavior was interpreted as a spin-flop transition. Since the Cr zig-zag chains form triangular Cr lattices, it was suggested that a competition between FM and AFM exchange interactions are present within the chains. This would lead to spin frustration (consistent with the supposed spin-flop transition) and a complex magnetic ordering in NaCr$_2$O$_4$, similar to the isostructural compounds NaV$_2$O$_4$ and CaCr$_2$O$_4$ \cite{nozaki2010incommensurate,damay2010zigzag}). The fact that the negative CMR in this material only appears below $T_{\rm N}$, suggests that this effect should be closely related to the magnetic structure. Therefore, spin frustration was regarded as a main responsible for the presence of unconventional CMR in NaCr$_2$O$_4$.

Earlier magnetic measurements using muon spin rotation/relaxation ($\mu^+$SR) and neutron powder diffraction (NPD) methods of the NaCr$_2$O$_4$ compound have actually been published \cite{nozaki,nozaki2014internal,jun}. Such studies revealed the occurrence of a commensurate canted antiferromagnetic (C-AFM) order with a Néel temperature $T_{\rm N}=125$~ K. A spin structure was presented with the Cr moments aligned ferromagnetically along the $b-$axis in each chain, and antiferromagnetically in the $ac-$plane among adjacent chains, with propagation vector $k$~=~(1~0~1) \cite{nozaki}. However, the value of the canting angle could not be precisely estimated. 
In this work we revisit this interesting compound using high-quality samples studied by state-of-the-art $\mu^+$SR and high resolution NPD techniques. We present a detailed and thorough study of the temperature dependent magnetic properties in NaCr$_2$O$_4$ that shed new light on the underlying physics and reveal a series of completely novel properties that were previously overlooked or misinterpreted 
\footnote{
The discrepancy between the present and previous studies of this compound is most likely related to a problem with the sample presented in especially Ref.~\cite{nozaki}. The size of the main magnetic peak and thereby also the ordered magnetic moment is clearly smaller in Ref.~\cite{nozaki} than in our present study. We have carefully confirmed our present results using several different sample batches as well as using a multitude of $\mu^+$SR and NPD instruments to ensure the robustness of our current conclusions. In fact the NPD data presented in Ref.~\cite{nozaki} is most likely instead representative of a sample with significant Ca doping that suppresses the ordered Cr moment \cite{Nocerino_NaCa}.
}.
Among other things, the canting angle of the Cr spin axial vector and its magnitude was clarified in detail. Moreover, an additional incommensurate (IC) modulation of the magnetic structure was observed. A slightly revised mechanism underlying the magnetic ordering was proposed and indications of a tricritical behavior as well as, possibly, metamagnetism emerged from the temperature dependence of the magnetic order parameter. Our results are consistent with the presence of high- and low-spin Cr sites in the magnetic ground state of NaCr$_2$O$_4$. Such Cr spin states are also foreseen by a recently published theoretical investigation, according to which a spin disproportionation exists between the Cr1 and Cr2 sites in NaCr$_2$O$_4$ \cite{yamaoka2020electronic}. The intra-chain FM alignment of the Cr moments along the $b-$axis, and a weak coupling between the magnetic ordering and the crystal structure, suggest that no spin frustration is in place along the zigzag chains. In fact, in other Q1D calcium-ferrite type compounds the magnetic order develops along the geometrically frustrated Cr zigzag ladders (e.g., $\beta$-CaCr$_2$O$_4$ and NaMn$_2$O$_4$ \cite{damay2010zigzag,matsubara2020neutron}). However, the magnetic interactions in NaCr$_2$O$_4$ instead "escape" the Q1D Cr chains and realize a 2D mangetic structure, developing across the double rutile strings in the $ac-$plane. This observation is also consistent with electronic structure calculations \cite{toriyama2014electronic}, which suggest that such kind of 2D magnetic structure is a very likely ground state. The absence of geometrical frustration, and the indications of tricritical behavior of the magnetic order parameter, suggest that the spin flop transition observed as a function of external magnetic field \cite{sakurai2012unconventional} could instead be interpreted as metamagnetism. 

\section*{Results}

In the following section the experimental results with the related data analysis are collected. In the first subsection the temperature dependent $\mu^+$SR measurements are presented, while the second subsection shows the NPD measurements for different temperatures.

\subsection*{\label{nopres}$\mu^+$SR Results}
\begin{figure}[ht]
  \begin{center}
    \includegraphics[scale=0.57]{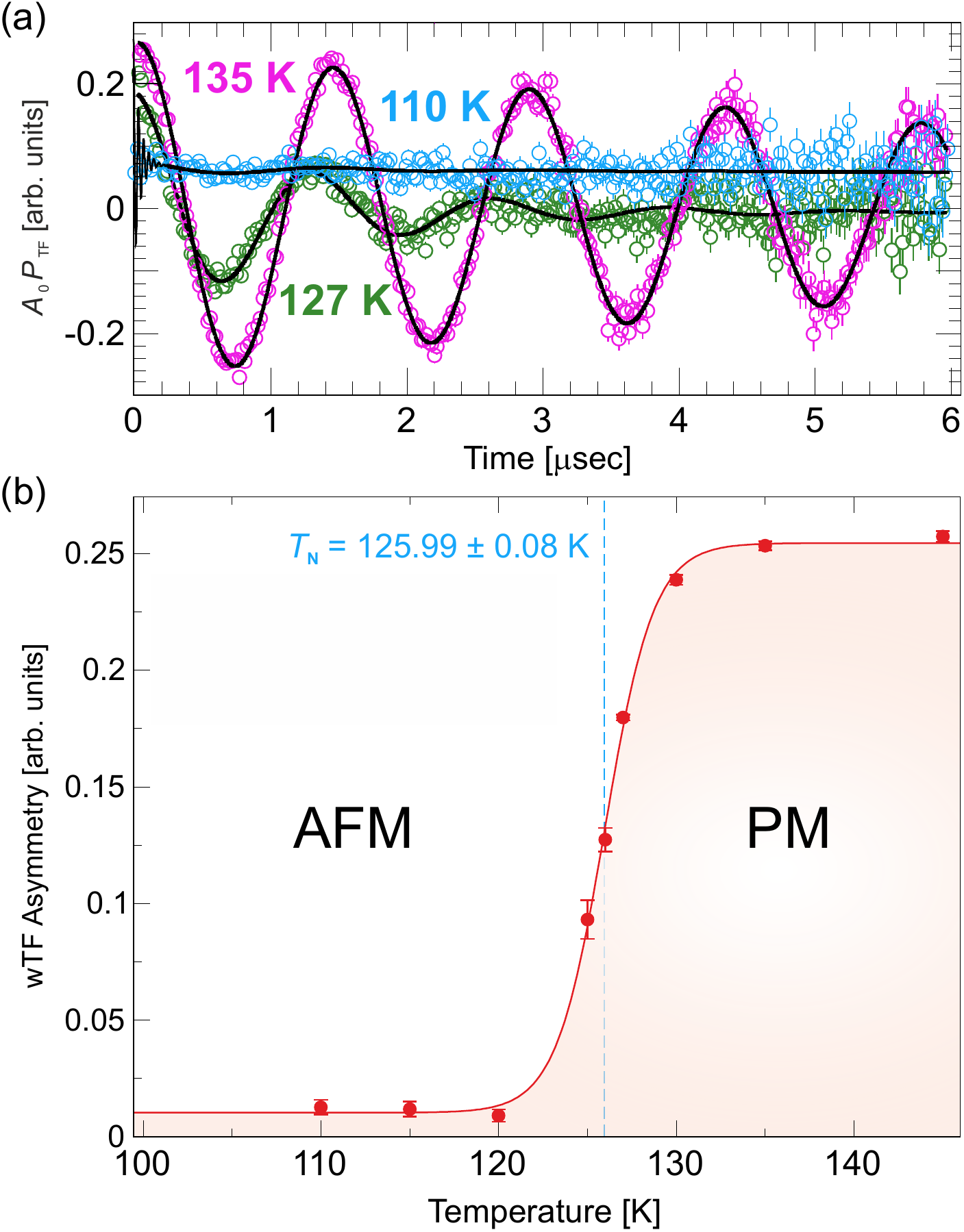}
  \end{center}
  \caption{(a) Weak transverse field (wTF) $\mu^+$SR spectra in the long time domain above the transition temperature for NaCr$_2$O$_4$. The spectra are overlapped to show the oscillation around zero asymmetry. (b) The corresponding wTF asymmetry is plotted as a function of temperature, with a fit to a sigmoid curve shown as a solid line. The transition temperature $T_{\rm N}\approx126$~K resulting from the fit is indicated by the vertical dashed line.}
  \label{allasy}
\end{figure}

In the wTF $\mu^+$SR measurements an external magnetic field is applied orthogonal to the initial spin-polarization of the muon beam, causing it to precess around the external field´s direction. In this configuration, $\mu^+$SR time spectra have been acquired in the temperature range $T=2-145$~K. The applied field wTF = 50G, is several orders of magnitude smaller than the internal field at the muon sites. Selected wTF spectra acquired for the NaCr$_2$O$_4$ are shown in Fig.~\ref{allasy}(a).
At high temperature the magnetic moments in the sample are randomly oriented, therefore the local internal magnetic field is negligible with respect to the external field, which causes the muon spins to precess accordingly. Indeed the spectrum at $T=145$~K exhibits a very regular oscillation, with a single frequency related to the wTF. As the temperature decreases, the system gradually goes from the paramagnetic (PM) to the AFM phase. The strong internal AFM field overcomes the weak external field, the overall depolarization rate increases and the amplitude of the wTF oscillation in the spectra are reduced, until only the high frequency oscillations in the early time domain are left [Fig.~\ref{allasy}(a)].
The fit function chosen for the wTF spectra is the following:
\begin{eqnarray}
 A_0 \, P_{\rm TF}(t) &=& A_{\rm TF}\cos(2\pi \nu_{\rm TF}t + \frac{\pi \phi}{180})\cdot{}e^{(-\lambda_{\rm TF} t)}
+ A_{\rm AFM}\cos(2\pi \nu_{\rm AFM}t + \frac{\pi \phi}{180})\cdot{}e^{(-\lambda_{\rm AFM} t)} + A_{\rm tail}\cdot{}e^{(-\lambda_{\rm tail} t)}.
\label{4}
\end{eqnarray}
Here $A_0$ is the initial asymmetry, $P_{\rm TF}$(t) is the muon spin polarization function, $A_{\rm TF}$, $A_{\rm AFM}$ and $A_{\rm tail}$ are the asymmetries of the related polarization components, $\nu_{\rm{TF/AFM}}$ is the frequency of the Larmor precession (whose value is related to the applied wTF and to the internal AFM field respectively), $\phi$ is the initial phase of the oscillating signal (here zero), $\lambda_{\rm TF}$, $\lambda_{\rm AFM}$ and $\lambda_{\rm tail}$ are the depolarisation rates. The magnetic transition temperature can be determined by fitting the temperature dependence of the transverse field asymmetry $A_{\rm TF}$ to a sigmoid function. Figure~\ref{allasy}(b) shows a plot of the wTF asymmetries as a function of temperature. Here an increase in $A_{\rm TF}$ is observed as across the magnetic phase transition where the system evolves from the low temperature AFM state to the PM state above $T_{\rm N}$. The middle point of the sigmoid fitting curve provides the magnetic transition temperature $T_{\rm N}=(125.99\pm0.08)$~K.

Inside the AFM phase (e.g., at $T=110$~K), the low frequency associated with the wTF is completely suppressed, as clearly seen in the raw data in Fig.\ref{allasy} and the signal remains unchanged down to $T=2$~K (therefore the $A_{\rm TF}$ temperature points below 110 K are not shown in the plot). The complete suppression of wTF asymmetry means that the entire sample volume enters the AFM phase. Above $T_{\rm N}$, the wTF asymmetry fully recovers its maximum value, implying that the muons in the entire volume of the sample are precessing according to the weak external field. The smooth transition from full asymmetry to complete suppression of $A_{\rm TF}$, without intermediate steps down to base temperature, rules out the possibility of a phase separation and/or substantial magnetic impurities.

\begin{figure*}[ht]
  \begin{center}
    \includegraphics[keepaspectratio=true,width=130 mm]{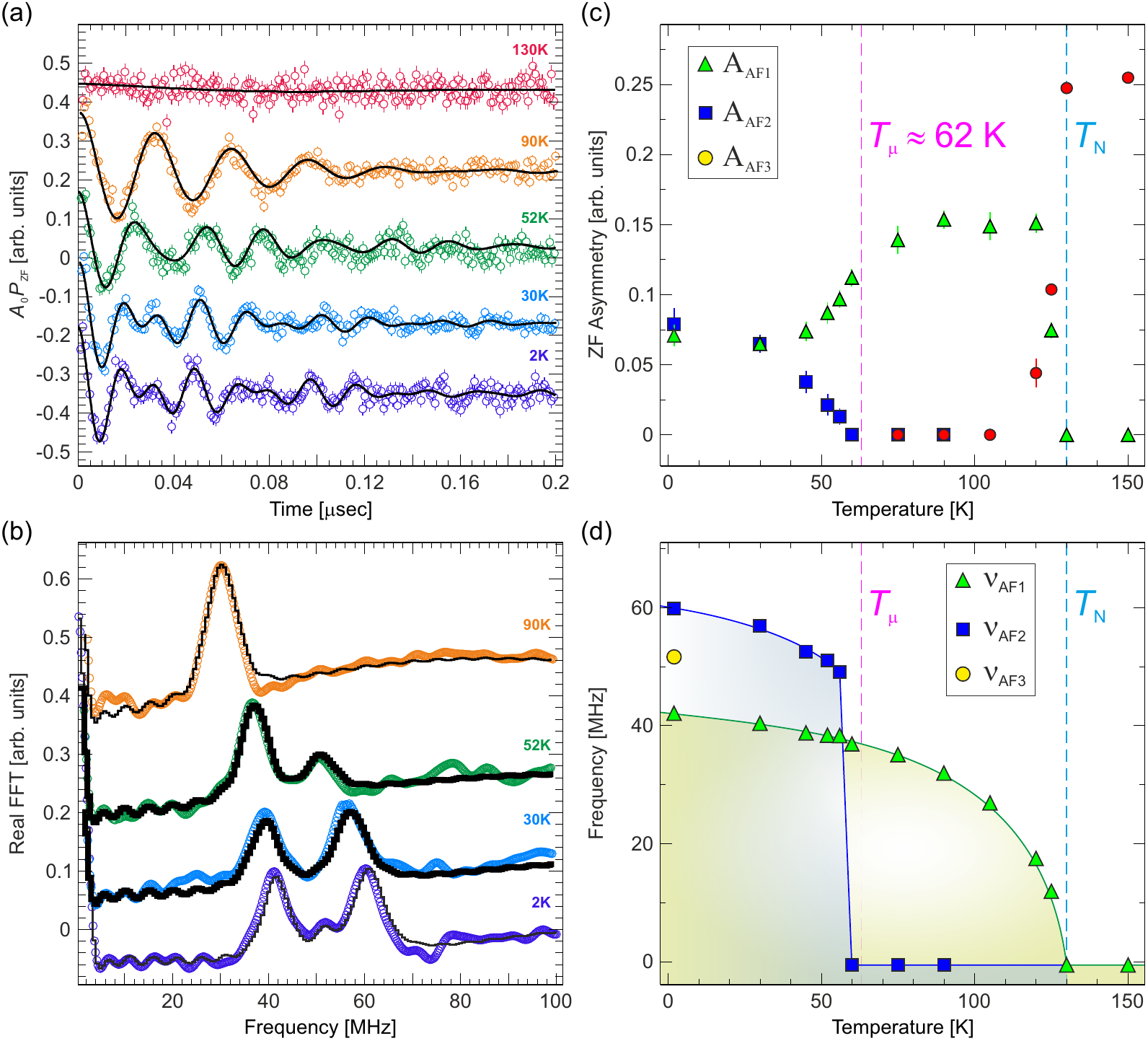}
  \end{center}
  \caption{(a) Early time domain $\mu^+$SR spectra in zero field (ZF) as a function of temperature across the magnetic phase transition revealing an evolution from three to one frequency. The continuous black lines are best fits to the function shown in Eq.~\ref{3}. For clarity of display, the spectra have been shifted along the $y-$axis. (b) Real part of the fast Fourier transform (FFT) for the NaCr$_2$O$_4$ ZF time spectra displayed in (a), with the continuous black lines again representing fits to Eq.~\ref{3}. (c) Temperature dependencies of the asymmetries for the oscillating components of the ZF signal. (d) Temperature dependence of the frequencies of the muon spin Larmor precession. Vertical dashed lines indicate critical temperatures, $T_{\mu}\approx62$~K and $T_{\rm N}\approx130$~K, respectively.}
  \label{ZFvsfr}
\end{figure*}

A zero-field (ZF) $\mu^+$SR experiment has also been performed for different temperatures, in order to observe the evolution of the internal magnetic field distribution in NaCr$_2$O$_4$. The muons implanted in the sample act as ultra sensitive microscopic magnetometers, which probe the internal field by following their spin polarization \cite{blundell1999spin}. The resulting ZF $\mu^+$SR time spectra for different temperatures in the short time domain are displayed in Fig.~\ref{ZFvsfr}(a).
Here, the clear oscillations visible in the spectra are linked to the muons spins' Larmor precession, whose angular frequency ($\omega$) changes according to the local magnetic field (of modulus B). The time dependence of the muon spin polarization can be described by exponentially relaxing oscillating functions. At base temperature $T=2$~K the fit function chosen for NaCr$_2$O$_4$ is the following:
\begin{eqnarray}
 A_0 \, P_{\rm ZF}(t) &=& \sum_{i = 1, 3}
A_{\rm AFi}\cos(2\pi \nu_{\rm AFi}t + \frac{\pi \phi_{\rm AFi}}{180})\cdot{}e^{(-\lambda_{\rm AFi} t)} + A_{\rm tail}\cdot{}e^{(-\lambda_{\rm tail} t)}.
\label{3}
\end{eqnarray}
where $A_0$ is the initial asymmetry of the muon decay, $P_{\rm ZF}$ is the muon spin polarization function. Here the distinct static internal magnetic field at the muon sites, associated to the commensurate AFM ordering, results in an oscillation of the muon spin polarization (fitted with a cosine function) with three well defined frequencies for the Larmor precession: $\nu_{\rm AF1}$ = (41.68 $\pm$ 0.24) MHz, $\nu_{\rm AF2}$ = (60.95 $\pm$ 0.31) MHz and $\nu_{\rm AF3}$ = (51.80 $\pm$ 0.31) MHz. $\phi$ is the initial phase-shift of the implanted muon polarization with respect to the detectors' plane. For NaCr$_2$O$_4$ we find that $\phi\approx0$ for all frequencies across the entire temperature range. Further, $\lambda_{\rm AF}$ is the exponential relaxation rate of the cosine function, and $A_{\rm tail} \cdot e^{(-\lambda_{\rm tail} t)}$ is a non oscillatory exponential tail term due to the components of the internal field that are parallel to the initial polarization direction of the muons' spins. This term is commonly found in powder samples. The presence of two clear frequencies is visible in the 2 K muon time spectra and its Fourier transform in Fig.~\ref{ZFvsfr}(a,b). At this temperature a third frequency could also be included in the fit, however, its small amplitude (compared to the other two frequencies) makes its presence in the muon spectrum questionable.
The asymmetries and frequencies of the ZF spectra as a function of temperature are displayed in Fig.~\ref{ZFvsfr}(c,d). As the temperature increases the spontaneous oscillation of the muon spin polarization, is suppressed until only the tail component dominates the spectrum above $T_{\rm N}$. In addition, the second frequency $\nu_{\rm AF2}$ undergoes is fully suppressed for temperatures above $T_{\mu}\approx62$~K [Fig.~\ref{ZFvsfr}(d)]. Such a drastic change in the frequencies are usually related to either a structural phase transition (changing the muon stopping sites) or a change in the spin structure (e.g. spin canting). However, as we will show below in our detailed NPD study, there is no change in the atomic or magnetic structure occurring in this temperature range. Looking more carefully at the ZF asymmetries in Fig.~\ref{ZFvsfr}(c), we see that with increasing temperature, there is a gradual shift of asymmetry from $A_{\rm AF2}$ to $A_{\rm AF1}$. Finally, when crossing $T_{\mu}$, only $A_{\rm AF1}$ (i.e. $\nu_{\rm AF2}$) survives. Most likely this means that two muon stopping sites are favorable below $T_{\mu}$, but with increasing temperature the second stopping site becomes energetically unfavorable. More detailed computer modeling of the muon stopping sites in NaCr$_2$O$_4$ will be required to confirm such hypothesis.

\begin{figure*}[ht]
  \begin{center}
    \includegraphics[scale=0.7]{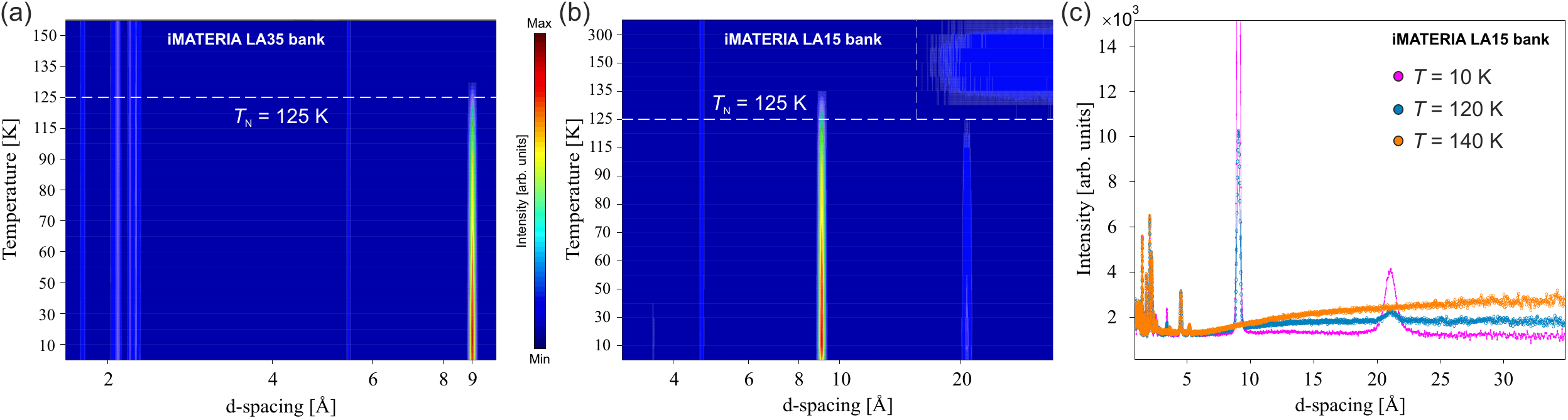}
  \end{center}
  \caption{(a) 2D color map of the temperature dependence of the NaCr$_2$O$_4$ diffraction pattern acquired in the iMATERIA low-angle bank (LA35) shown in $d$-spacing. A large magnetic peak is clearly visible at $d=9$~Å, and the transition temperature $T_{\rm N}=125$~K is shown as horizontal dashed line. (b) 2D color map of the temperature dependence of the diffraction pattern in NaCr$_2$O$_4$ acquired using the iMATERIA low(er)-angle bank (LA15) shown in $d$-spacing. An additional, weaker, magnetic peak is visible at at $d\approx20$~Å. Above $T_{\rm N}$, a broad diffuse scattering intensity is clearly visible in the $d$-range surrounding such weak magnetic peak, as indicated by the dashed square region. (c) Diffraction patterns from the LA15 iMATERIA bank at different temperatures, showing the evolution of the magnetic diffuse scattering, forming the magnetic Bragg peak below $T_{\rm N}$. Please note that the $y-$axes for both 2D color maps in (a) and (b) panels have non-linear temperature scales.}
  \label{color_plot}
\end{figure*}

\subsection*{NPD Results}
To study the temperature evolution of the magnetic structure in NaCr$_2$O$_4$, neutron powder diffraction (NPD) patterns were collected at the time-of-flight (ToF) instrument iMATERIA in J-PARC while comparative low temperature patterns were collected also using the SPICA instrument. Figures~\ref{color_plot}(a-c) show the temperature evolution of the diffraction pattern in the high-$d$ (low-$Q$) range. Here, the absence of splitting for the nuclear Bragg peaks in the low-$d$ range, indicates that no structural transitions occur in the investigated temperature range [Fig.~\ref{color_plot}(a)]. A high-intensity magnetic peak (and several other), appearing at $T_{\rm N}=125$~K, is clearly seen at $d=9$~Å, as expected from previous characterizations. In addition, a weaker magnetic peak appears at even higher $d\sim20$~Å for $T<T_{\rm N}$. These two magnetic reflections manifest a slightly different phenomenology. Indeed, in the $d$-range around 20 Å, a broad diffuse scattering intensity is clearly visible above $T_{\rm N}$ [inside the dashed rectangle in Fig.~\ref{color_plot}(b)], while no such scattered intensity can be observed around $d$ = 9 Å.
The diffuse scattering indicates the presence of short-range spin correlations in the temperature range between 130 K and 300 K. Below 130 K, such disordered state is replaced by a long range magnetic order, and the  diffuse scattered intensity condenses into a Bragg peak at $d$ $\sim$ 20 Å [Fig. \ref{color_plot}(c)]. The presence of high-temperature short range magnetic correlations in NaCr$_2$O$_4$ was never reported before. The behavioral difference between the strong and the weak magnetic peaks suggests that they are originated by two distinct periodic arrangements of the magnetic moments in the Cr lattice. 

\begin{figure*}[ht]
  \begin{center}
    \includegraphics[scale=0.95]{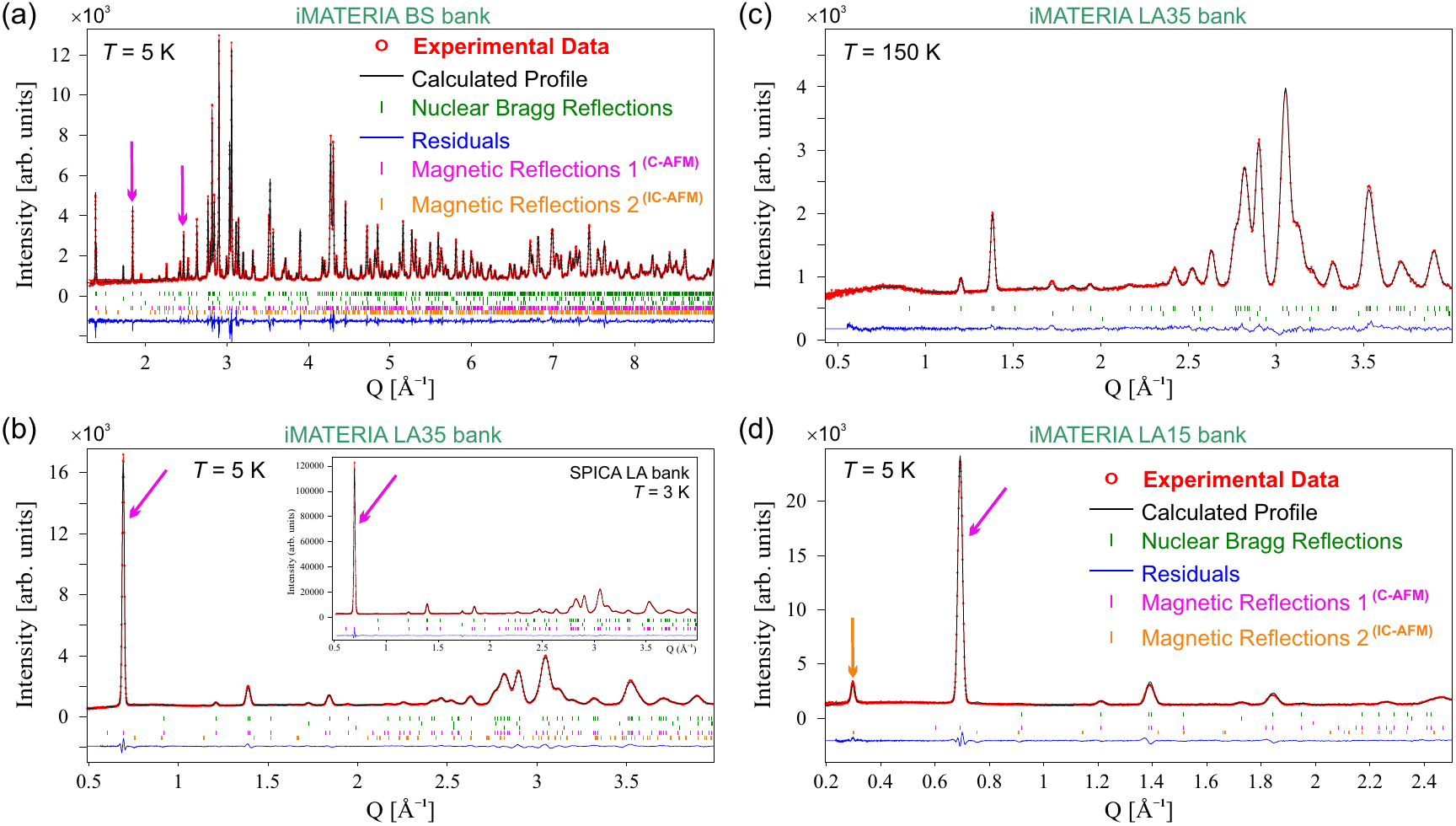}
  \end{center}
  \caption{Neutron diffraction pattern of NaCr$_2$O$_4$ acquired from the (a) the backscattering (BS) and (b) 35$^{\circ}$ (LA35) low-angle banks of iMATERIA at base temperature. The inset in (b) shows the diffraction pattern of NaCr$_2$O$_4$ from the low-angle LA bank of SPICA at base temperature (for comparison). (c) Neutron diffraction pattern from the 35$^{\circ}$ (LA35) bank of iMATERIA, above the transition temperature (no magnetic peaks). (d) Neutron diffraction pattern from the 15$^{\circ}$ (LA15) bank of iMATERIA, at base temperature. The arrows indicate the magnetic Bragg peaks in the diffraction patterns, where magenta and orange colors refer to the commensurate (C-AFM) and incommensurate (IC-AFM) magnetic propagation vectors, respectively.}
  \label{neut}
\end{figure*}

The magnetic and crystal structure of NaCr$_2$O$_4$ has been determined with a global procedure involving the simultaneous Rietveld refinement of diffraction patterns from all 4 ToF detector banks while keeping the cell parameters, the atomic positions and the basis vectors of the magnetic moment as common parameters. A comparative refinement for data collected in one of the low angle banks from SPICA at $T=3$~K was used as additional confirmation and provided very similar results. 
The low temperature refined profiles with the experimental data from both instruments measured with different detector banks are displayed in Fig.~\ref{neut}(a-d), where the magnetic peaks are highlighted by magenta and orange arrows.
The observed profiles are in very good agreement with the calculated models, as also underlined by the low values of the reliability R-factors, reported in Table~\ref{tabstruc} along with refined structural parameters at both high and low temperature.

\begin{table*}[ht]
\renewcommand{\arraystretch}{1.25}
\small
  \caption{\ Structural lattice parameters, atomic positions and isotropic thermal displacement parameters ($B_{\rm iso}$) for NaCr$_2$O$_4$, as refined from neutron powder diffraction data (iMATERIA, global fit to all detector banks). Also shown are the refinement reliability factors for the nuclear [$R_{\rm B(nuc)}$] and magnetic [$R_{\rm B(mag1)}$ and $R_{\rm B(mag2)}$] phases. All the atomic positions have the site symmetry $4c$.}
  \label{tabstruc}
  \small
  \begin{tabular*}{0.75\textwidth}{@{\extracolsep{\fill}}  c| c c}
      \hline
                        & $T=300$~K        & $T=5$~K       \\
    \hline

    Space Group         & $Pnma$  & $Pnma$  \\
    $a$ (Å)            & 9.01556(5)       & 9.0305(1)       \\
    $b$ (Å)            & 2.91279(1)       & 2.91695(3)      \\
    $c$ (Å)            & 10.41412(6)       & 10.3428(1)       \\
    $\alpha$ = $\beta$ = $\gamma$      & 90$^{\circ}$     & 90$^{\circ}$       \\

    \hline

    Na (x, y, z)    &  (0.243(2), 1/4, 0.347(1)) &  (0.244(2), 1/4, 0.348(1)) \\
    Cr1 (x, y, z)    &  (0.082(1), 1/4, 0.603(1)) &  (0.083(2), 1/4, 0.603(1)) \\
    Cr2 (x, y, z)    &  (0.063(1), 1/4, 0.114(1)) &  (0.062(2), 1/4, 0.115(1)) \\
    O1 (x, y, z)    & (0.2958(8), 1/4, 0.6489(7))  &  (0.297(1), 1/4, 0.6499(8)) \\
    O2 (x, y, z)    & (0.3860(8), 1/4, 0.9803(7))  &  (0.3867(9), 1/4, 0.9805(9)) \\
    O3 (x, y, z)    & (0.4768(9), 1/4, 0.2170(7))  &  (0.476(1), 1/4, 0.2177(9)) \\
    O4 (x, y, z)    & (0.0756(9), 1/4, 0.9295(8))  &  (0.074(1), 1/4, 0.9287(8)) \\

    \hline

    $B_{\rm iso}$(Na)  (Å$^2$) & 0.503(5)  & 0.39(2)  \\
    $B_{\rm iso}$(Cr1)  (Å$^2$) & 0.058(9)  &  0.025(8) \\
    $B_{\rm iso}$(Cr2)  (Å$^2$) & 0.093(3)  &  0.18(2) \\
    $B_{\rm iso}$(O1)  (Å$^2$) &  0.121(1)  &  0.09(1) \\
    $B_{\rm iso}$(O2)  (Å$^2$) & 0.165(1)  &  0.14(2) \\
    $B_{\rm iso}$(O3)  (Å$^2$) & 0.147(1)  &  0.23(1) \\
    $B_{\rm iso}$(O4)  (Å$^2$) & 0.207(1)  &  0.19(2) \\

    \hline

    $R_{\rm B(nuc)}$ (\%)   & 3.64         & 2.98         \\
    $R_{\rm B(mag1)}$ (\%)   & -         & 0.802         \\
    $R_{\rm B(mag2)}$ (\%)   & -         & 7.14         \\
    \hline
  \end{tabular*}
\end{table*}

\begin{figure*}[ht]
  \begin{center}
    \includegraphics[scale=0.8]{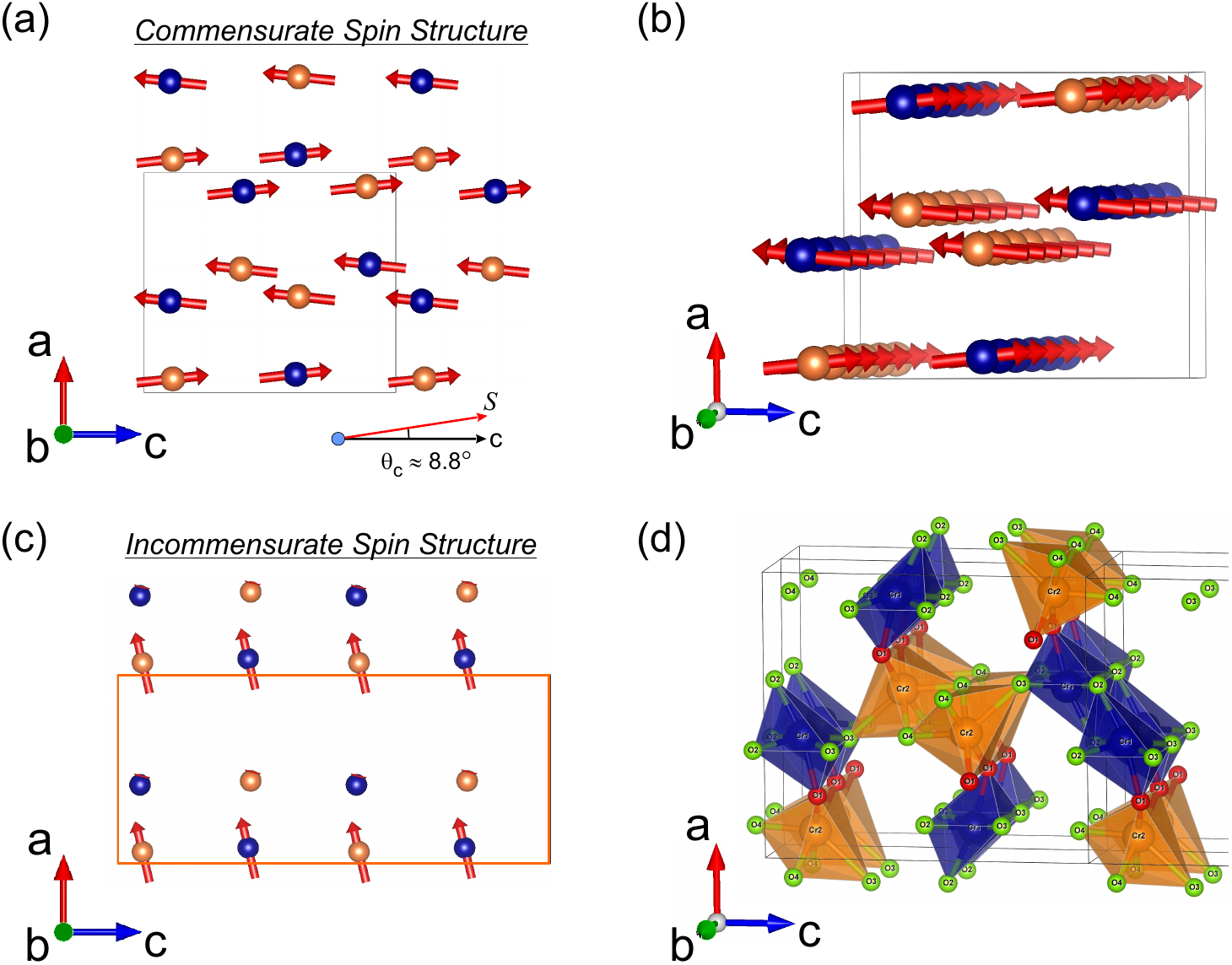}
  \end{center}
  \caption{(a,b) Commensurate (C-AFM) spin structure of NaCr$_2$O$_4$ with $k_1$~=~(1~0~1) at $T=5$~K. The Na atoms have been removed for clarity of display. The spin canting [$\theta_{\rm c}\approx(8.8\pm0.5)^{\circ}$] and the AFM coupling of the chains along the $a-$axis is visible from the $ac-$plane orientation. The ferromagnetic alignment of the Cr spin within the chains along the $b-$axis is emphasized in the (b) panel. (c) Incommensurate (IC-AFM) magnetic supercell with $k_2$~=~(0~0~$\sim$1/2). The spin alignment in this case is cycloidal, but a systematic suppression of the anti-parallel Cr moments along the $a-$axis also occurs. (d) CrO$_6$ octahedra with the oxygen sites explicitly labelled. The oxygen site (O1) that mediates the AFM coupling between the zig-zag Cr chains is highlighted in red.}
  \label{spin}
\end{figure*}

The calculated diffraction pattern contains a nuclear and a two magnetic NaCr$_2$O$_4$ phases, plus two small impurity Cr$_2$O$_3$ and CrO$_2$ impurity phases. The high and low temperature nuclear phase was found to be well modeled by a orthorhombic $Pnma$ space group (No. 62), in agreement with previously reported results. In our high-resolution data we also found that the adoption of strain parameters (only relevant for the high resolution BS bank pattern) substantially improved the refinement in the full temperature range. The strain is probably a heritage of the high pressure synthesis process necessary to manufacture the sample. The Rietveld refinement of the NaCr$_2$O$_4$ magnetic phases includes the two magnetic Cr atoms in their respective crystallographic sites. The scale factors and structural parameters were constrained to be equal to their counterparts in the nuclear NaCr$_2$O$_4$ phase, for a proper estimation of the Cr moments $\mu_{Cr}$ in the two magnetic cells. The indexing of the temperature dependent magnetic Bragg reflections was achieved with the magnetic propagation vectors $k_1$~=~(1~0~1) and $k_2$~=~(0~0~0.497)~=~(0~0~${\textstyle \frac{1}{2}-}\delta$), determined with the software K-Search. While the $k_1$ magnetic cell is commensurate (C-AFM) to the crystal structure, the $k_2$ cell is incommensurate (IC-AFM) and has the $c-$axis doubled with respect to the nuclear and $k_1$ magnetic cells (see also Fig.~\ref{spin}).

The software BasIreps provided the possible irreducible representations of the propagation vector group $G_{\rm k}$ compatible with the space group $Pnma$ and the propagation vectors $k_1$ and $k_2$. The direct sum of 8 possible irreducible representations (IRrep) gave the reducible representation for $k_1$ as \textbf{$\Gamma$} = 1$\Gamma_1$ $\bigoplus$ 2$\Gamma_2$ $\bigoplus$ 2$\Gamma_3$ $\bigoplus$ 1$\Gamma_4$ $\bigoplus$ 1$\Gamma_5$ $\bigoplus$ 2$\Gamma_6$ $\bigoplus$ 2$\Gamma_7$ $\bigoplus$ 1$\Gamma_8$. Among the different IRreps used to refine the magnetic phase with a trial and error approach, the $\Gamma_7$ of dimension 1 contained 2 times in \textbf{$\Gamma$} was selected as the one that could provide the best match. For $k_2$ the reducible representation can be written as \textbf{$\Gamma$} = 1$\Gamma_1$ $\bigoplus$ 2$\Gamma_2$ $\bigoplus$ 1$\Gamma_3$ $\bigoplus$ 2$\Gamma_4$. Among the different IRreps used to refine the magnetic phase with a trial and error approach, the $\Gamma_4$ of dimension 1 contained 2 times in \textbf{$\Gamma$} was selected as the one that could provide the best match. Table~\ref{table1} reports the real and imaginary basis vectors of the $\Gamma_7$ ($k_1$) and $\Gamma_4$ ($k_2$) representation for the crystallographic symmetry operators associated with the two Cr sites (Cr1 and Cr2).

\begin{table}[h]
\renewcommand{\arraystretch}{1.25}
\small
  \caption{\ Basis vectors of the irreducible representations $\Gamma_7$ ($k_1$) and $\Gamma_4$ ($k_2$) of the magnetic propagation vector group G$_k$.}
  \label{table1}
  \begin{tabular*}{0.48\textwidth}{@{\extracolsep{\fill}}  c c c}
    \hline
    & $\Gamma_7$ ($k_1$): C-AFM & \\
    \hline
    SYMM         & x,y,z &   \\
    BsV$_1$(Re)    & (1 0 0)  & (0 0 1)  \\
    BsV$_1$(Im)    & (0 0 0)  & (0 0 0) \\
    \hline
    SYMM          & -x+1/2,-y,z+1/2 &   \\
    BsV$_2$(Re)    & (1 0 0)  &  (0 0 -1)   \\
    BsV$_2$(Im)    & (0 0 0)  & (0 0 0) \\
    \hline
    SYMM         & -x,y+1/2,-z &   \\
    BsV$_3$(Re)    & (1 0 0)  & (0 0 1)  \\
    BsV$_3$(Im)    & (0 0 0)  & (0 0 0) \\
    \hline
    SYMM         & x+1/2,-y+1/2,-z+1/2 &   \\
    BsV$_4$(Re)    & (1 0 0)  &  (0 0 -1)   \\
    BsV$_4$(Im)    & (0 0 0)  & (0 0 0)) \\
    \hline
    \hline
       & $\Gamma_4$ ($k_2$): IC-AFM & \\
   \hline
    SYMM           & x,y,z &   \\
    BsV$_1$(Re)    & (1 0 0)  & (0 0 1)  \\
    BsV$_1$(Im)    & (0 0 0)  & (0 0 0) \\
    \hline
    SYMM           & -x+1/2,-y,z+1/2 &   \\
    BsV$_2$(Re)    & (-0.022 0 0)  &  (0 0 0.022)   \\
    BsV$_2$(Im)    & (-1 0 0)  & (0 0 1) \\
    \hline
  \end{tabular*}
\end{table}

The coefficients of these basis vectors for both $k_1$ and $k_2$, namely C1 and C2, were refined as common parameters for both the Cr sites, since no significant improvement for the refinement was observed by leaving them independent. The resulting spin axial vector for $k_1$ has a large component along the crystallographic $c-$direction, a smaller component along the $a-$direction and zero component along the $b-$direction, leading to a tilt of the Cr spin with respect to the $c-$axis. Due to the presence of imaginary basis vectors for the IRrep of $k_2$, the resulting spin axial vector has a cycloidal-like modulation along the $a-$axis, reminiscent of the magnetic structure found in the isostructural compound NaMn$_2$O$_4$ \cite{matsubara2020neutron}. The latter material also exhibits two distinct C-AFM and IC-AFM magnetic cells. However, while in NaMn$_2$O$_4$ the AFM coupling and the IC modulation of the Mn moments develop along the zigzag Mn chains, in NaCr$_2$O$_4$ the magnetic structures develop along the rutile chains in the $ac-$plane. As a result, the C-AFM alignment of the Cr spins exhibits canting, and the IC cycloidal Cr spin arrangement display systematic suppression of the Cr moment on the spin inversion sites. A graphic representation of the obtained magnetic structures is displayed in Fig.~\ref{spin}(a-c). For the $k_1$ cell, the canted AFM coupled FM double rutile chain along the $c-$axis, are clearly visible in Fig.~\ref{spin}(a), while the FM zig-zag chains running along the $b-$axis are emphasized in Fig.~\ref{spin}(b). For the $k_2$ cell, the modulated cycloidal arrangement of the Cr moments along the $a-$axis is displayed in Fig.~\ref{spin}(c). Figure~\ref{spin}(d) explicitly shows the different Cr and O sites in the lattice. including the oxygen site (O1) that mediates the AFM coupling between the zig-zag Cr chains.

For the $k_1$ structure, the canting angle $\theta_{\rm c}=(8.8\pm0.5)^{\circ}$, estimated from the average value of the components of the basis vector C1 and C2 through the temperature range from base-T to 100K. This angle is estimated with higher confidence with respect to the one estimated previously \cite{nozaki}. The total moment of the Cr atoms, calculated as the modulus of the Cr spin axial vector averaging among the low temperature fluctuations, is $\mu^{\rm C}_{\rm Cr}=(4.30\pm0.01)\mu_{\rm B}$ for the $k_1$ structure, and $\mu^{\rm IC}_{\rm Cr}=(2.20\pm0.03)\mu_{\rm B}$ for the $k_2$ structure.

\section*{Discussion}

The $\mu^{\rm C}_{\rm Cr}$ value found in this work is remarkably higher than the expected spin only value for a system with a nominal valence state Cr$^{3.5+}$, since $S_{\rm Cr^{3+}}$ = 3.87 $\mu_{\rm B}$ and S$_{\rm Cr^{4+}}$ = 2.83 $\mu_{\rm B}$ (which are actually consistent with $\mu^{\rm IC}_{\rm Cr}$). Values closer to the measured $\mu^{\rm C}_{\rm Cr}$ value could be achieved by including the orbital contributions to the total Cr moment, which would sum up to an ordered moment of 5.2 and 4.5 $\mu_{\rm B}$ for Cr$^{3+}$ and Cr$^{4+}$, respectively. However, orbital contributions from Cr in an octahedral environment is expected to give negligible contribution to the total moment. This is due to orbital quenching in a crystal field \cite{datta1934xlix}, making this option less probable. Indeed, even though orbital contributions of vanadium in octrahedral environment was in fact recently suggested in VI$_3$ by $\mu^+$SR and XAS \cite{forslund2022refined,de2022influence}, for NaCr$_2$O$_4$, an unusual mixed valance electronic state has instead been reported \cite{taguchi2017unusual}. 
It was suggested that Na substitution in the solid solution Na$_x$Ca$_{1-x}$Cr$_2$O$_4$ does not result in electrons removed from the Cr site, as expected in typical mixed valence compounds. Here, the electrons are instead removed from the O sites, to relax charge frustration and realize a negative charge transfer state \cite{taguchi2017unusual}. In fact, the large value of the Cr moment ($\mu^{\rm C}_{\rm Cr}$) measured in this work could be explained by the negative charge transfer state induced in NaCr$_2$O$_4$, since it could lead to a Cr valence Cr$^{2+}$ in a high spin state $d^3$ ($t^3_{2g}$ $e^1_g$). The expected spin only value for Cr$^{2+}$ is 4.9 $\mu$B, which is comparable to $\mu^{\rm C}_{\rm Cr}$. Therefore, our results could indirectly confirms the unusual charge transfer present in NaCr$_2$O$_4$ \cite{taguchi2017unusual}. While it was not commented on, since CO is not observed in this system \cite{toriyama2014electronic,takeda2013magnetic}, this kind of electronic configuration also suggest highly fluctuating vacancies. A similar situation was also observed for the 1D mixed valance system K$_2$Cr$_8$O$_{16}$ \cite{bhobe2015electronic}, which is structurally analogous to NaCr$_2$O$_4$. 

Additionally, our result is consistent with the predictions of recently published DFT calculations \cite{yamaoka2020electronic}, which envision a spin disproportionation between the Cr1 and Cr2 sites. This would lead to the existence of high-spin Cr sites in the magnetic ground state of NaCr$_2$O$_4$. The authors of Ref.~\cite{yamaoka2020electronic} evaluated the modulus of the Cr moment as $\approx2\mu_{\rm B}$ for both low and high spin states, which is close to the value found in Ref.~\cite{nozaki}. Nevertheless, they also acknowledge that, due to the uncertainty of the results of Ref.~\cite{nozaki}, their estimation cannot be experimentally validated.
The large value of $\mu^{\rm C}_{\rm Cr}$ obtained in this work is also consistent with a magnetic coupling mechanism mediated by ligand holes in NaCr$_2$O$_4$, as also suggested in references \cite{sakurai,taguchi2017unusual}. This aspect will be further expanded below in the discussion related to the coupling mechanism involved in the magnetic order. 

\begin{figure*}[ht]
  \begin{center}
    \includegraphics[scale=0.9]{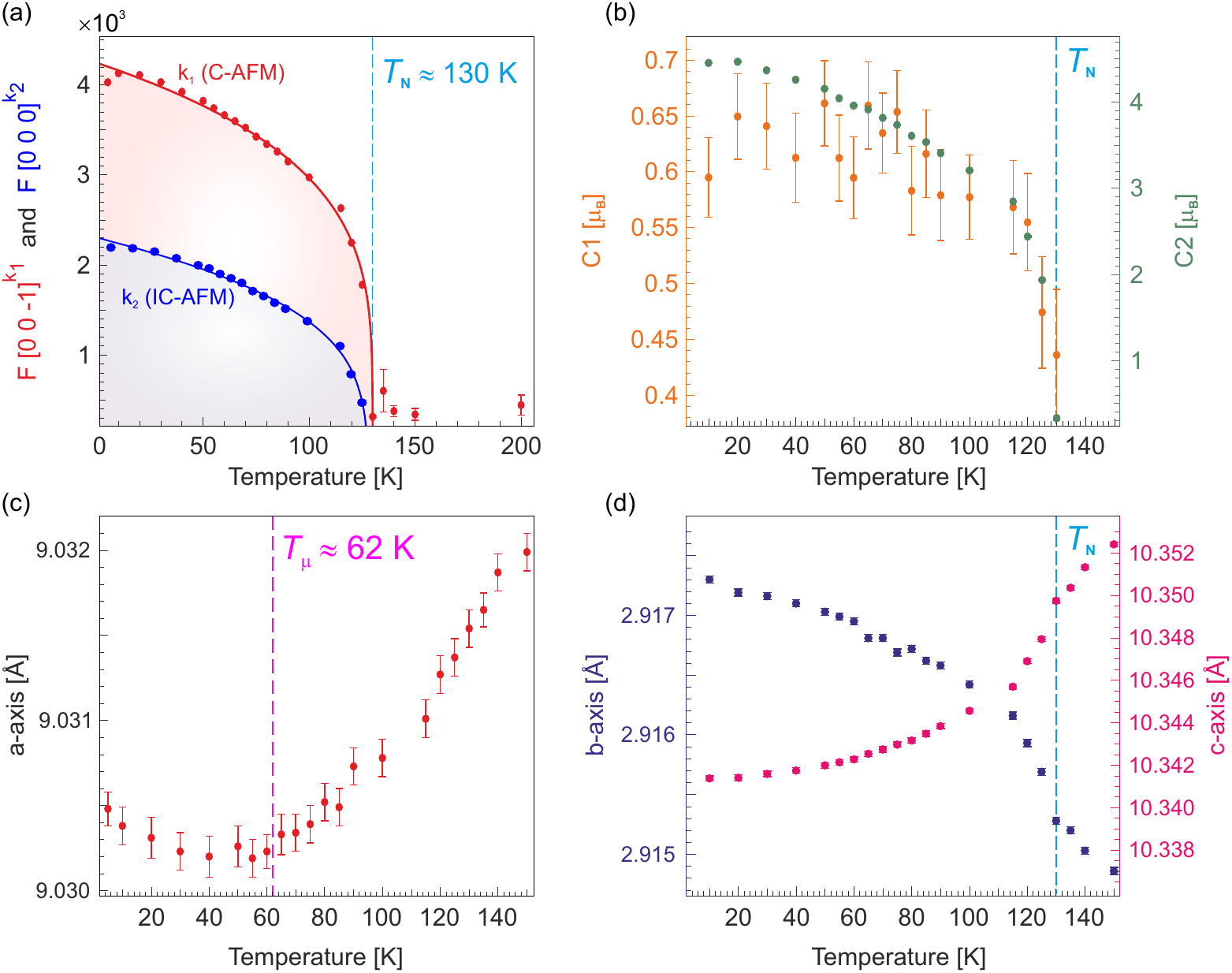}
  \end{center}
  \caption{(a) Structure factor ($F$) of the large $k_1$ magnetic peak (0~0~-1) as well as the weaker $k_2$ magnetic peak (0~0~0) with fits to Eq.~\ref{bcs} shown as a solid line. (b) Temperature dependence of the coefficients of the $k_1$ basis vectors, C1 and C2. (c-d) Temperature dependence of the unit cell parameters. The magnetic transition temperature ($T_{\rm N}$) along with the anomaly related to changes in the number of muon frequencies ($T_{\mu}$, see also Fig.~\ref{ZFvsfr}) are shown as vertical dashed lines.}
  \label{struc_factor}
\end{figure*}

The temperature dependence of the structure factor F[0 0 -1], obtained as the square root of the integrated intensity of the of the large $k_1$ magnetic peak (0 0 -1) and smaller $k_2$ magnetic peak (0 0 0), are plotted in Fig.~\ref{struc_factor}(a). The solid lines are fits to the power law:
\begin{eqnarray}
f(x) \propto (1 - \frac{x}{T_c})^{\beta}.
\label{bcs}
\end{eqnarray}

The temperature evolution of the diffraction peak is related to the order parameter of the magnetic phase transition. By fitting it with Eq.~\ref{bcs} it is possible to estimate the onset of the magnetic transition temperature, and acquire some indications about the type of symmetry breaking that the system undergoes. The onset of the magnetic transition was found at $T=(130\pm0.003)$~K, in good agreement with the $\mu^+$SR measurements. The value of the critical exponent, estimated for the $k_1$ peak, was found $\beta_1=0.245\pm 0.007\approx\frac{1}{4}$, suggesting that the AFM-PM phase transition in NaCr$_2$O$_4$ is located at a tricritical point \cite{yaouanc2011muon}. Within the Landau mean-field approximation, this means that the onset of the magnetic transition in NaCr$_2$O$_4$ is a point between the edge of a second order transition and the ordinary critical point of a first-order transition \cite{huang2008statistical,griffiths1970thermodynamics}. The tricritical behavior observed in this work, together with the dramatic increase in the magnetization with the relatively small externally applied field of 3.5 T (associated to a spin-flop transition) \cite{sakurai2012unconventional}, are compatible with the fact that a possible hidden metamagnetic state \cite{stryjewski1977metamagnetism} is established in this material at the edge of the AFM-PM transition. This conjecture is based on the qualitative agreement between NaCr$_2$O$_4$ and tricritical metamagnetic systems that manifest a similar phenomenology, i.e., NiCl$_2$, 2H$_2$O, Ni(NO$_3$)$_2$ \cite{swuste1977magnetic,schmidt1970metamagnetism,polgar1971low}. In these systems, magnetic field dependent changes in magnetization and susceptibility were erroneously identified as spin-flop transitions. Nevertheless, studies of the evolution of the magnetic structure under high magnetic fields would be necessary to resolve the ambiguity of the field induced transitions, and to outline of the magnetic phase diagram of NaCr$_2$O$_4$.
The temperature dependencies of both the coefficients of the $k_1$ basis vectors C1, C2 follow the same trend as the structure factor F[0 0 -1] [cf. Fig.~\ref{struc_factor}(a,b)], suggesting that the canting angle of the Cr spin axial vector is maintained across the full temperature range, as its modulus decreases. The temperature trend for the structure factor F[0 0 0] of the $k_2$ magnetic peak is also displayed in Fig.~\ref{struc_factor}(a). Here the critical exponent is estimated as $\beta_2=0.31\pm 0.01$, which is close to the value for a 3D Heisenberg AFM with a 1-dimensional order parameter \cite{yaouanc2011muon}, suggesting a more conventional magnetic ground state for this phase.

The temperature dependence of the unit cell parameters are displayed in Fig.~\ref{struc_factor}(c,d). On cooling, the value of the $a-$axis decreases monotonically down to $T\approx60$~K, where an inversion of trend is observed. Phenomenologically, such inversion coincides with the suppression of the second muon frequency at $T_{\mu}\approx62$~K. This could be a weak indication that the subtle changes in the lattice could yield an energetically unfavorable situation for one of the muon stopping sites. The value of the $b-$axis instead increases monotonically, with a trend similar to the one observed for the evolution of the magnetic parameters in Fig.~\ref{struc_factor}(a,b), with a clear change in slope at the magnetic transition temperature $T_{\rm N}=125$~K. The $c-$axis undergoes a monotonic decrease down to base temperature, with a similar but mirrored behavior as the one observed in the $b-$axis. The temperature evolution of the unit cell parameters shows evidence of coupling between the crystal and magnetic structure. However, since the magnetic ordering does not induce a structural transition towards a different lattice symmetry, but only very slight changes in the lattice parameters, such coupling does not seem to be very strong. Finally, the $\mu^+$SR precession signal displayed a gradual transition from two to one frequency at $T_{\mu}\approx62$~K [Fig.~\ref{ZFvsfr}(b)]. As already mentioned above, since no dramatic effect is present in the NPD patterns (neither nuclear nor magnetic) around these temperatures (Fig.~\ref{struc_factor}), the reason for the change in the frequency spectrum is probably due to a change in the structural muon sites. Here the temperature evolution of the lattice and atomic positions could simply result in that one of the muon stopping site becomes energetically unfavorable above $T_{\mu}$.

It appears that, depending on the reciprocal orientation of the octahedra, FM and AFM exchange interaction are established between the Cr ions. In particular, for the $k_1$ magnetic structure, three different magnetic couplings seem to be in place below $T_{\rm N}$: \textcircled{\small{1}} An inter-chain FM exchange mechanism is always established across the edge sharing octahedra in the Cr1 and Cr2 chains. \textcircled{\small{2}} An intra-chain FM exchange mechanism is always established across the corner sharing octahedra between the chains stacked along the $c-$direction (through the oxygen atom in the crystallographic site O3). \textcircled{\small{3}} an intra-chain AFM coupling is always established across corner sharing octahedra between the chains stacked along the $a$ direction (through the oxygen atom in the crystallographic site O1).
Consistently with the large value of the Cr moment extracted in this work, we can consider the ligand holes concentrated at the corner oxygen O1 and O3 sites \cite{toriyama2014electronic} as the main responsible for the occurrence of the C-AFM phase $k_1$. Using the arguments similar to the ones presented in reference \cite{korotin1998cro}, the FM inter-chain coupling can be indeed interpreted in terms of a double exchange FM mechanism mediated by the hopping of the O3 ligand holes. This would involve Cr orbitals with one localized electron per site and one with itinerant character, through the bond $\widehat{Cr1-O3-Cr2}$. In fact, the O ions on the edges shared by the octahedra in the zig-zag double chain (O2, O4), were found to give insignificant contributions to the bands near the Fermi level \cite{toriyama2014electronic}. Therefore, the observed inter-chain FM coupling can be considered a consequence of the intra-chain FM double exchange. This is compatible with the temperature evolution of the crystal structure, in which a monotonic increase of the $b-$axis on cooling suggests an increase of the Cr-Cr distances within the zig-zag double chain (i.e., compatible with a Coulomb repulsion effect). A similar phenomenology was observed in other systems such as K$_2$Cr$_8$O$_{16}$ and CrO$_2$ \cite{toriyama2011peierls,forslund2021neutron,korotin1998cro,schlottmann2003double}.

Despite the dominance of FM double-exchange interaction, intra-chain AFM coupling is established along the $c-$axis double rutile chains. It has been suggested that the AFM coupling is due to a weak super-exchange coupling through the bond $\widehat{Cr1-O1-Cr2}$ \cite{nozaki}. However, it can naively be argued that an AFM super-exchange interaction across a $\widehat{Cr-Anion-Cr}$ bond requires a bond angle of 180$^{\circ}$ \cite{kanamori1959superexchange}. Cases in which the AFM super-exchange mechanism occurs across bond angles that deviate from 180$^{\circ}$ can of course be realized in the presence of complex crystal-field symmetries of the magnetic ions (like in our case, with the distorted octahedral coordination of Cr ions). However, in NaCr$_2$O$_4$ the value of the bond angle $\widehat{Cr1-O1-Cr2}$ = 125.58(5)$^{\circ}$, is very far from 180$^{\circ}$ (is actually closer to 90$^{\circ}$, which would favor FM coupling not AFM). Therefore we suggest the mutual hopping of 2 itinerant electrons between the two Cr sites, aided by the O1 ligand holes, as a possible alternative mechanism to explain the intra-chain AFM coupling in NaCr$_2$O$_4$. The hopping of correlated itinerant electrons was suggested as the possible mechanisms to explain the magnetic ordering in CrO$_2$ \cite{korotin1998cro}. Such suggestion was though discarded since it would have lead to AFM ordering, which is not observed in CrO$_2$. Since the $\widehat{Cr1-O1-Cr2}$ and $\widehat{Cr1-O3-Cr2}$ bonds are structurally equivalent and the O1 and O3 sites are both preferred for the localization of the ligand holes, a similar hopping coupling mechanism would be expected. The only difference between the two bonds would be the mutual orientation of the electronic orbitals in the non-equivalent Cr sites. Here, the different distortions between Cr1O$_6$ and the Cr2O$_6$ octahedra, remove the degeneracy of the Cr-$3d$ orbitals, leading to one localized and two itinerant orbitals for the Cr1 site, and two localized and one itinerant orbital for the Cr2 site \cite{toriyama2014electronic}. Due to the lack of inversion symmetry between Cr ions, the canting of the moments in this spin configuration is consistent with an anisotropic AFM exchange including an antisymmetric Dzyaloshinsky-Moriya (DM) exchange term \cite{dzyaloshinsky1958thermodynamic,moriya1960anisotropic}.

The proposed correlated electron hopping mechanism for the AFM coupling is deduced as a logic consequence of these observations. In this regard, given the discrepancy between the moduli of the ordered moments of the two C-AFM and IC-AFM structures (i.e. $\mu^{\rm C}_{\rm Cr}=4.3~\mu_{\rm B}$ vs. $\mu^{\rm IC}_{\rm Cr}=2.2~\mu_{\rm B}$), and the different values of their critical exponents $\beta_1 \sim \frac{1}{4}$ and $\beta_2 \sim 0.31$, we suggest that the itinerant hopping electrons are responsible for the $k_1$ long-range C-AFM order, while the localized electrons are responsible for the IC-AFM order. To verify this conjecture, neutron diffraction studies under applied magnetic fields would be an optimal and direct test. Indeed, magnetic structures with itinerant character are sensitive to external magnetic fields, while the localized ones are not \cite{li2015distinct}.

To summarize, the detailed and partly revised magnetic structure of NaCr$_2$O$_4$ is presented in this work. The analysis of high-resolution data shed new light on this interesting compound. In addition to a canted C-AFM phase with propagation vector $k_1$~=~(1~0~1), an additional IC cycloidal-like magnetic phase is identified with propagation vector $k_2$~=~(0~0~${\textstyle \frac{1}{2}-}\delta$). Following the temperature evolution of the structure factor of the magnetic reflections, indications of unconventional critical behavior for the $k_1$ magnetic phase transition in NaCr$_2$O$_4$ were found. In particular, we suggest that the magnetic transition might be at the edge of a tricritical point with possible formation of a metamagnetic hidden phase. 
Weak interplay between the crystal and magnetic structure, evidenced by the comparison between the $\mu^+$SR and NPD data, is compatible with the absence of geometric frustration. The value of the ordered Cr moments of the C-AFM and IC-AFM structures were extracted as $\mu^{\rm C}_{\rm Cr}=4.3~\mu_{\rm B}$ and $\mu^{\rm IC}_{\rm Cr}=2.2~\mu_{\rm B}$, respectively. The high $\mu^{\rm C}_{\rm Cr}$ value is consistent with a double exchange coupling mechanism mediated by itinerant Cr moments (i.e., the ligand-hole hopping mechanism), for the $k_1$ C-AFM order. On the other hand, the lower $\mu^{\rm IC}_{\rm Cr}$ value is consistent with a coupling mechanism mediated by exchange interactions between localized Cr spins, for the $k_2$ IC order. The value of the canting angle of the Cr spin axial vector for the C-AFM structure has been determined with high accuracy as $\theta_c=(8.8 \pm0.5)^{\circ}$. A FM double-exchange interaction, together with an AFM correlated itinerant electrons hopping interaction, have been proposed as the mechanisms driving the magnetically ordered $k_1$ phase in NaCr$_2$O$_4$. 
Magnetic field dependent studies of the magnetic structure would be relevant to discern the effects of the localized and itinerant electrons on the ordered moments, and clarify the magnetic (H/T) phase diagram of the NaCr$_2$O$_4$ compound.

\section*{Methods}

Polycrystalline samples of NaCr$_2$O$_4$ were prepared from a stoichiometric mixture of NaCrO$_2$, Cr$_2$O$_3$, and CrO$_3$ at 1300$^{\circ}$C under a pressure of 7 GPa. The samples were synthesized at the National Institute for Material Science (NIMS) in Tsukuba, Japan. The samples were preliminary characterized using in-house x-ray diffraction (XRD). This showed that all sample batches were single phase, with a CaFe$_2$O$_4$-type $\textit{Pnma}$ structure. Further details on the sample preparation can be found in Ref.~\cite{sakurai}.

The $\mu^+$SR spectra have been acquired, as a function of temperature, at the multi purpose surface muon instrument Dolly \cite{dolly}, at the Paul Scherrer Institute (PSI) in Switzerland \cite{psi}. Approximately 500~mg of powder sample was packed in a 1$\times$1 cm$^2$ area envelope made from Aluminum-coated Mylar tape (0.05 mm thickness). The envelope was attached to a low background Cu sample holder inserted in a helium exchange gas cryostat (temperature range: 1.6 K to 300 K).

The neutron powder diffraction (NPD) measurements were performed at the time of flight powder diffractometers iMATERIA \cite{ishigaki2009ibaraki} and SPICA \cite{yonemura2014development} at the high intensity proton accelerator facility J-PARC, in Japan. The powder samples ($m\approx0.72$~g) were mounted into cylindrical vanadium cells with diameters 6 mm (for SPICA) and 5 mm (for iMATERIA). The cells were sealed using an aluminium cap, aluminium screws and and indium wire. The cell was mounted on a closed cycle refrigerator to reach temperatures between 2 K and 300 K. SPICA and iMATERIA are designed to reach a wide range in reciprocal space, while measuring at a single scattering angle by means of large position sensitive detectors. Different detector banks provide different d-ranges (Q-ranges) with gradually changing resolutions, in particular in SPICA the high angle detector bank, provides d-range from 0.3 Å up to 3.7 Å with a resolution $\Delta d / d$ = 0.12$\%$, while the low angle (LA) detector bank has a d-range from 0.5 Å up to 11 Å. In iMATERIA the backward detector bank (BS), allows a d-range from 0.181 Å up to 5.09 Å with a resolution $\Delta d / d$ = 0.16$\%$, while the low angle detector banks (LA35 and LA15) allow to resolve d-ranges from 0.25 Å up to 40 Å. These features make the high angle banks very suitable for detailed structural characterization while the low angle banks are ideal to unequivocally identify any magnetic Bragg peak, which usually appear in the high-d (low-Q) range. 

The crystal and magnetic structure determination was carried out with the FullProf software suite \cite{rodriguez1993recent}. All images involving crystal structure were made with the VESTA software \cite{momma}, the parameter fitting has been performed with the software Igorpro \cite{igor} and the $\mu^+$SR data were fitted using the \textit{musrfit} \cite{musrfit} software package.

\bibliography{Refs}



\section*{Acknowledgements (not compulsory)}

The $\mu^+$SR measurements were performed at instrument $Dolly$ of the Swiss Muon Source (S$\mu$S), at the Paul Scherrer Institute in Villigen, Switzerland. The NPD measurements were performed at the instruments iMATERIA and SPICA of the neutron spallation source J-PARC (beamtime proposals: 2019A0330, 2019B0420). The authors wish to thank the staff of PSI and J-PARC for their great support during the $\mu^+$SR and NPD experiments.
This research is funded by the Swedish Foundation for Strategic Research (SSF) within the Swedish national graduate school in neutron scattering (SwedNess), as well as the Swedish Research Council VR (Dnr. 2021-06157 and Dnr. 2017-05078), and the Carl Tryggers Foundation for Scientific Research (CTS-18:272). J.S. is supported by the Japan Society for the Promotion Science (JSPS) KAKENHI Grant No. JP18H01863 and JP20K21149. Y.S. and O.K.F. are funded by the Chalmers Area of Advance - Materials Science.

\section*{Author contributions statement}

E.N., J.S., and M.M. conceived the experiments. E.N., O.K.F., N.M., F.M., D.A., J-C.O., T.S, I.U., Y.S., J.S. and M.M. conducted the experiments. E.N,  O.K.F., J.S., N.M., and M.M. analyzed the results. The samples were synthesized by H.S., who also conducted the initial sample characterizations. E.N. and M.M made all the figures. E.N. created the first draft, and all co-authors reviewed and revised the manuscript. 

\section*{Data availability statement}

All the data of this work are available from the corresponding authors on request.

\textbf{Competing interests} 

The authors declare no competing interests.  

The corresponding author is responsible for submitting a \href{http://www.nature.com/srep/policies/index.html#competing}{competing interests statement} on behalf of all authors of the paper. 

\end{document}